# Spin-dependent transport properties in GaMnAs-based spin hot-carrier transistors

Yosuke Mizuno

*Department of Electronic Engineering, The University of Tokyo, 7-3-1 Hongo, Bunkyo-ku, Tokyo 113-8656, Japan*

Shinobu Ohya

*Department of Electronic Engineering, The University of Tokyo, 7-3-1 Hongo, Bunkyo-ku, Tokyo 113-8656, Japan, and PRESTO, Japan Science and Technology Agency, 4-1-8 Honcho, Kawaguchi-shi, Saitama 332-0012, Japan*

Pham Nam Hai

*Department of Electronic Engineering, The University of Tokyo, 7-3-1 Hongo, Bunkyo-ku, Tokyo 113-8656, Japan*

Masaaki Tanaka

*Department of Electronic Engineering, The University of Tokyo, 7-3-1 Hongo, Bunkyo-ku, Tokyo 113-8656, Japan, and SORST, Japan Science and Technology Agency, 4-1-8 Honcho, Kawaguchi-shi, Saitama 332-0012, Japan*

We have investigated the spin-dependent transport properties of GaMnAs-based 'three-terminal' semiconductor spin hot-carrier transistor (SSHCT) structures. The emitter-base bias voltage $V_{EB}$ dependence of the collector current $I_C$, emitter current $I_E$, and base current $I_B$ shows that the current transfer ratio $\alpha$ (= $I_C / I_E$) and the current gain $\beta$ (= $I_C / I_B$) are 0.8-0.95 and 1-10, respectively, which means that GaMnAs-based SSHCTs have current amplifiability. In addition, we observed an oscillatory behavior of the tunneling magnetoresistance (TMR) ratio with the increasing bias, which can be explained by the resonant tunneling effect in the GaMnAs quantum well.

III-V-based ferromagnetic-semiconductor heterostructures containing GaMnAs can be a good model system for future spintronic devices. Large tunneling magnetoresistance (TMR) of 75% (8.0 K) [1] and 290% (0.39 K)[2] were observed in the GaMnAs-based single-barrier magnetic tunnel junctions (MTJs). Also, the resonant tunneling effect was observed in the GaMnAs quantum-well heterostructures, which indicates that highly coherent tunneling occurs in the GaMnAs-based heterostructures.[3] However, there are no reports concerning ferromagnetic-semiconductor-based 'three-terminal' spin devices with amplification capability. In three-terminal devices, the spin-dependent transport properties can be changed by bias voltages. If amplifiability is attached to the devices, they can be applied to integrated circuits with novel functionalities. As such spin devices, magnetic tunnel transistors (MTTs) have recently captured a lot of attention.[4-6] Metal-based MTTs are composed of ferromagnet (FM) / insulator (I) / FM / semiconductor (SC), whose output currents can be controlled by the bias voltages and magnetization orientation of the FM layers, but it is very difficult to fabricate metal-based MTTs with amplifiability due to the frequent scatterings in the metallic base. Here, we propose and fabricate a semiconductor spin hot-carrier transistor (SSHCT). SSHCT is structurally similar to conventional metal-based MTTs but composed of all-epitaxial semiconductor heterostructures, which enables advanced band engineering and has good compatibility with the existing semiconductor technology. We have recently carried out a numerical simulation of the device characteristics of SSHCTs, and showed that, unlike metal-based MTTs, SSHCTs potentially have amplifiability and that we can control the sign of their magnetocurrent (MC) ratio due to the resonant tunneling effect.[7] By taking advantage of these features, SSHCTs are expected to have application possibilities not merely for non-volatile memories and high-accuracy magnetic-field sensors but also for the components of multi-valued logic and reconfigurable logic circuits.[8] In this letter, we report on the spin-dependent transport properties of GaMnAs-based 'three-terminal' SSHCT structures. Extremely high transfer ratio $\alpha$ and current gain $\beta$ were obtained, indicating that GaMnAs-based SSHCTs have current amplifiability. The oscillatory behavior of the bias dependence of the TMR ratio was observed, which was successfully explained by the resonant tunneling effect in the GaMnAs quantum well.

We grew SSHCT structures composed of (from the top surface to the substrate) $Ga_{0.95}Mn_{0.05}As$ (30 nm)/ GaAs (1 nm)/ AlAs (2 nm)/ GaAs (1 nm)/ $Ga_{0.95}Mn_{0.05}As$ (30 nm)/ Be-doped GaAs (30 nm) on p-type GaAs(001) substrates using low-temperature molecular-beam epitaxy (LT-MBE). The Be concentration of the Be-doped GaAs (GaAs:Be) layer was $1\times10^{17} cm^{-3}$. Figure 1(a) shows their layered structure, the substrate temperature, and the reflection high-energy electron diffraction (RHEED) pattern during the growth process. The two 1-nm-thick GaAs spacer layers are inserted to prevent the Mn diffusion into the AlAs layer and to obtain atomically flat interfaces. The $1\times2$ streaky RHEED pattern indicates that the crystal quality of the GaMnAs layers is extremely high. The p-type GaAs(001) substrates were backed with thick In layers to obtain ohmic contacts. Figure 1(b) shows the schematic valence band diagram of the GaMnAs-based SSHCT structures. In this diagram, spin-polarized hot holes are injected from the GaMnAs emitter into the GaMnAs base by tunneling. During the transmission through the ferromagnetic GaMnAs base, they lose energy due to spin-dependent scattering. Only those carriers that maintain enough energy to go over the Schottky-like barrier at the base/collector interface can contribute to the collector current. The SSHCT structures were fabricated into three-terminal devices with an emitter (E), a base (B), and a collector (C), as schematically shown in Fig. 1(c), by standard photolithography and chemical wet etching techniques. Etching was stopped in the base layer, on which a metal electrode was formed by evaporating Au. The active area of the tunnel junction and the total area of the base layer are $50 \times 100$ μm$^2$ and $100 \times 200$ μm$^2$, respectively. The formation of the device structure was confirmed both by a topology measurement using an atomic force microscope (AFM) and by a current-voltage (*I-V*) measurement between each pair of the three terminals (E-B, E-C, and B-C). The *I-V* characteristics of E-B and E-C showed a tunneling behavior, and those of B-C showed a Schottky behavior. The following transport measurements were performed at 2.6 K with a conventional two-terminal direct-current (DC) method in common-base configuration. The temperature 2.6 K is sufficiently lower than the Curie temperatures $T_C$ of GaMnAs in the emitter and the base layers, which are roughly estimated to be 40 K by the temperature dependence of the resistance between the emitter and the base terminals.

Figure 2 shows the transistor characteristics, *i.e.*, the $V_{BC}$ dependence of the collector current $I_C$ with various $V_{EB}$ ranging from 0 to 1000 mV with 100 mV steps. With increasing $V_{EB}$, the energy of the hot holes injected into the base layer becomes higher, which results in the increase of the number of those carriers that can go over the collector barrier. Meanwhile, with increasing $V_{BC}$, the effective thickness of the Schottky-like collector barrier is reduced, which promotes the tunneling phenomena through the barrier. In this way, $I_C$ is well controlled both by $V_{BC}$ and $V_{EB}$.

Figure 3 shows the $V_{EB}$ dependence of $I_C$, $I_E$, and $I_B$ when $V_{BC}$ = 0.1 mV. The current transfer ratio $\alpha$ (= $I_C / I_E$) is estimated to be 0.8-0.95, which is much higher than the maximum value ($\alpha$ = 0.03) reported in metal-based MTTs.[6] The inset of Fig. 3 shows the current gain $\beta$ (= $I_C / I_B$) *vs.* $V_{EB}$. The noise at low $V_{EB}$ (0.04 V) is due to the detection limit of our measuring equipment. The obtained $\beta$ increases linearly with $V_{EB}$, and is of the order of 10, which means that GaMnAs-based SSHCTs have current amplifiability. In a simple model, $\alpha$ is given by $\alpha = \exp(-t/\lambda)$, where $t$ is the base layer thickness, and $\lambda$ is the hole energy attenuation length. In general, metal-based MTTs are thought to have $\lambda$ of less than 3 nm.[9,10] However, GaMnAs-based SSHCTs are considered to have $\lambda$ of several tens of nanometers,[3] since they are composed of fully epitaxial single crystal layers, which leads to the drastic enhancement of $\alpha$ and $\beta$.

The inset of Fig. 4(a) shows the TMR curve when the bias voltage $V_{EC}$ between the emitter and the collector is 350 mV at 2.6 K with a magnetic field ***B*** applied along the [$\bar{1}10$] axis in plane. Here, *RA* in the vertical axis is the resistance area, (tunnel resistance $V_{EC} / I_C$) × (junction area). The base electrode was kept open. Clear TMR (maximum 5.3%) was observed. In this way, $I_C$ is well controlled by the parallel and antiparallel magnetization orientation of the emitter and the base layers. Figure 4(a) shows the $V_{EC}$ dependence of the TMR ratio at 2.6 K when a magnetic field ***B*** is applied along the [$\bar{1}10$] axis in plane. With increasing $V_{EC}$, the TMR ratio dropped sharply at around $V_{EC}$ = 77 mV. Then it reached its maximum 5.3% when $V_{EC}$ is 350 mV, and monotonically decreased after that. Even when $V_{EC}$ = 2 V, clear TMR (less than 0.1%) still remained. This bias dependence of the TMR ratio had good reproducibility in different devices. On the grounds that TMR disappears at some hundreds of mV in normal GaMnAs-based single-barrier MTJs,[11,12] the actual voltage applied between the emitter and the collector is assumed to be approximately one tenth of $V_{EC}$, because a large part of $V_{EC}$ contributes only to the deformation of the Schottky-like collector barrier. This assumption is appropriate considering that $V_{half}$, at which TMR is reduced by half, is about 450 mV (= 800 − 350, as shown in Fig. 4(a)), which is ten times larger than those of GaMnAs-based single-barrier MTJs.[2,11,12] The peak of the TMR ratio at $V_{EC}$ = 350 mV thus corresponds to the TMR maximum occasionally observed at some tens of mV in GaMnAs-based MTJs,[3] which can be reproduced by a theoretical calculation based on the transfer-matrix method and the Esaki-Tsu formula.[13]

Figure 4(b) shows the $V_{EC}$ dependence of $RA$ between the emitter and the collector in parallel and antiparallel configuration. In Fig. 4(b), the two curves have a kink at around $V_{EC}$ = 77 mV, where the TMR ratio sharply drops. The magnified view at the kink is shown in the inset of Fig. 4(b). Both curves have local minimal values of $RA$, but at slightly different $V_{EC}$. The local minima are considered to result from the resonant tunneling effect. The quantum level (77 / 10 = 7.7 meV) is close to the calculation result of 10 meV of the first quantum level of heavy holes. We can estimate the spin-splitting energy $\Delta E_{HH}$ of the first quantum level of heavy holes by the difference of the voltages $V_{EC}$ where $RA$ becomes minimal in parallel and antiparallel configuration. Since the difference of $V_{EC}$ is 1 mV (= 77 − 76), $\Delta E_{HH}$ is calculated to be 0.1 meV (= 1 / 10), which is consistent with the calculation result that $\Delta E_{HH}$ is negligible small in GaMnAs layers when magnetization is in plane.[14] Since the heavy hole spins are oriented along the tunneling direction and the *p-d* exchange Hamiltonian[15] is proportional to **s** · **S**, where **s** and **S** are spins of the carrier and the Mn atom, respectively, the quantum levels of heavy holes are not spin split but those of light holes are split by the in-plane magnetization.

In conclusion, we have measured the spin-dependent transport properties of GaMnAs-based 'three-terminal' SSHCT structures. Extremely high transfer ratio *α* and current gain *β* were obtained, indicating that GaMnAs-based SSHCTs have current amplifiability. The oscillatory behavior of the bias dependence of the TMR ratio was explained by the resonant tunneling effect in the GaMnAs quantum well.

This work was partly supported by PRESTO / SORST of JST, Grant-in-Aids for Scientific Research, IT Program of RR2002 of MEXT, and Kurata-Memorial Hitachi Science & Technology Foundation.

**Figure Captions**

Fig 1. (a) Schematic layered structure of GaMnAs-based semiconductor spin hot-carrier transistors (SSHCTs) consisting of $Ga_{0.95}Mn_{0.05}As$ (30 nm)/ GaAs (1 nm)/ AlAs (2 nm)/ GaAs (1 nm)/ $Ga_{0.95}Mn_{0.05}As$ (30 nm)/ GaAs:Be (30 nm; doping density $N_A = 1 \times 10^{17} cm^{-3}$) grown on *p*-type GaAs(001) substrates, accompanied by substrate temperature $T_s$ and reflection high energy electron diffraction (RHEED) patterns during the growth process. (b) Schematic valence band diagram of GaMnAs-based SSHCT structures. Region 1 is the emitter, region 2 is the AlAs tunnel barrier, and region 3 is the base. The collector is the *p*-type GaAs(001) substrate (region 4). (c) Schematic device structure of the GaMnAs-based SSHCTs fabricated in this study.

Fig 2. Transistor characteristics, *i.e.*, the collector current $I_C$ as a function of $V_{BC}$ at 2.6 K with various $V_{EB}$ ranging from 0 to 1000 mV (each step is 100 mV).

Fig 3. Collector current $I_C$, emitter current $I_E$, and base current $I_B$ as a function of $V_{EB}$ at 2.6 K at $V_{BC} = 0.1$ mV. The inset shows the current gain $\beta$ (= $I_C / I_B$) as a function of $V_{EB}$.

Fig 4. (a) TMR ratio as a function of $V_{EC}$ at 2.6 K when a magnetic field ***B*** is applied along the $[\bar{1}10]$ axis in plane. The inset shows the TMR curve; the resistance area $RA$ = (tunnel resistance $V_{EC} / I_C$) × (junction area) as a function of magnetic field $B$ at 2.6 K when $V_{EC}$ = 350 mV. (b) $RA$ as a function of magnetic field $B$ when the field is applied along the $[\bar{1}10]$ axis in plane at 2.6 K. The inset shows its magnified view at the kink around $V_{EC} = 77$ mV.

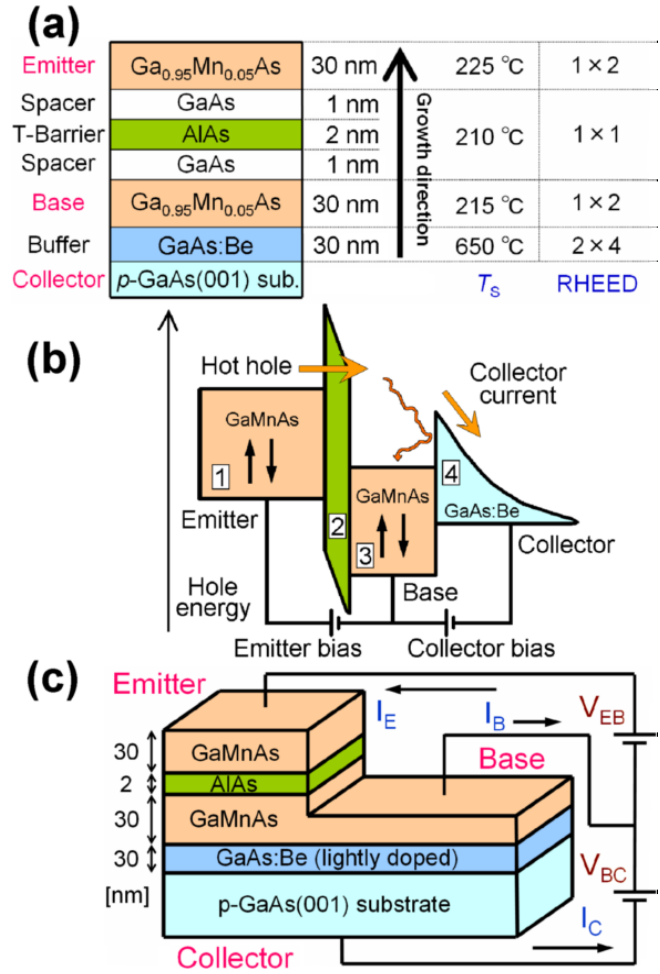

Fig. 1

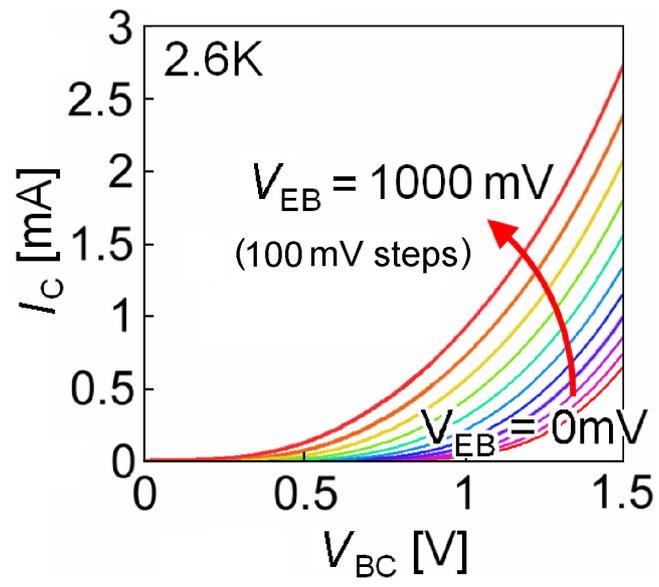

Fig. 2

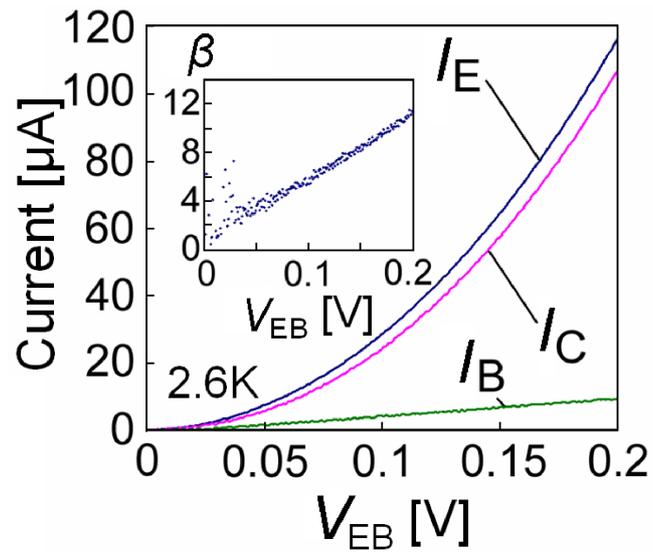

Fig. 3

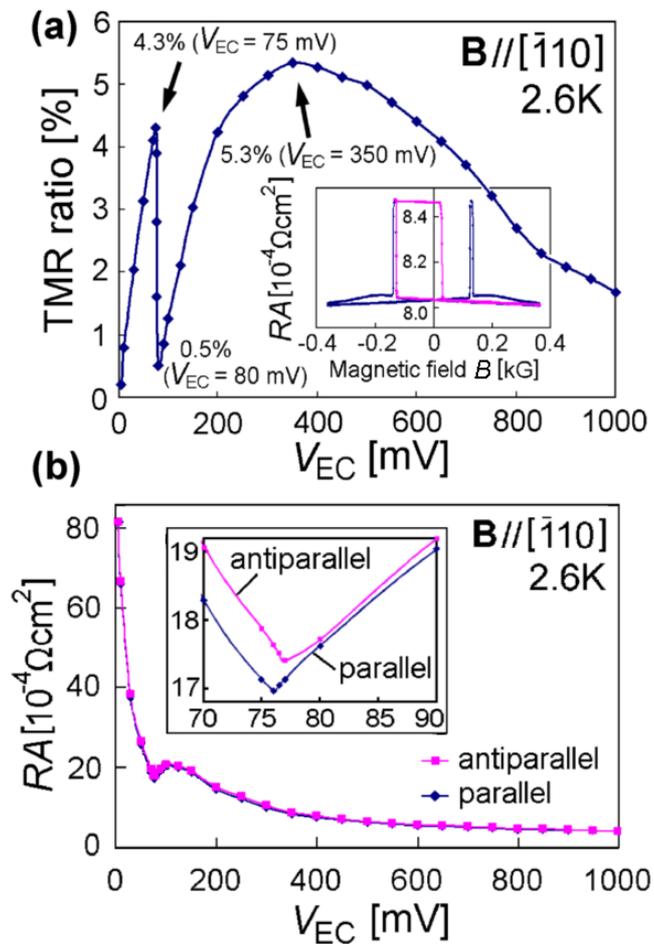

Fig. 4